\documentclass[]{aastex701}
\usepackage{amsmath}
\usepackage{amssymb}
\usepackage{graphicx}
\usepackage{xcolor}
\usepackage{listings}
\usepackage{hyperref}
\usepackage{subcaption}
\usepackage{comment}
\usepackage{url}
\usepackage{natbib}

\graphicspath{{./}{Figure/}}

\begin{document}

\title{Development of an early warning method incorporating pre-supernova neutrino light curves}

\author{K.Saito}
\affiliation{Research Center for Neutrino Science, Tohoku University, Sendai 980-8578, Japan}
\email{saito@awa.tohoku.ac.jp}
\author{M.Eizuka}
\affiliation{Research Center for Neutrino Science, Tohoku University, Sendai 980-8578, Japan}
\email{}
\author{Z.Hu}
\affiliation{Department of Physics, Kyoto University, Kyoto, Kyoto 606-8502, Japan}
\affiliation{Ecole Polytechnique, IN2P3-CNRS, Laboratoire Leprince-Ringuet, F-91120 Palaiseau, France}
\email{}
\author{K.Ichimura}
\affiliation{Research Center for Neutrino Science, Tohoku University, Sendai 980-8578, Japan}
\email{}
\author{M.Ikeda}
\affiliation{Kamioka Observatory, Institute for Cosmic Ray Research, University of Tokyo, Kamioka, Gifu 506-1205, Japan}
\affiliation{Kavli Institute for the Physics and Mathematics of the Universe (WPI), The University of Tokyo Institutes for Advanced Study, University of Tokyo, Kashiwa, Chiba 277-8583, Japan}
\email{}
\author{K.Ishidoshiro}
\affiliation{Research Center for Neutrino Science, Tohoku University, Sendai 980-8578, Japan}
\email{}
\author{N.Kawada}
\affiliation{Research Center for Neutrino Science, Tohoku University, Sendai 980-8578, Japan}
\email{}
\author{L.N.Machado}
\affiliation{School of Physics and Astronomy, University of Glasgow, Glasgow, Scotland, G12 8QQ, United Kingdom}
\email{}
\author{Ll.Marti}
\affiliation{Kavli Institute for the Physics and Mathematics of the Universe (WPI), The University of Tokyo Institutes for Advanced Study, University of Tokyo, Kashiwa, Chiba 277-8583, Japan}
\email{}
\author{K.Mikami}
\affiliation{Research Center for Neutrino Science, Tohoku University, Sendai 980-8578, Japan}
\email{}
\author{K.Tachibana}
\affiliation{Research Center for Neutrino Science, Tohoku University, Sendai 980-8578, Japan}
\email{}
\author{R.A.Wendell}
\affiliation{Department of Physics, Kyoto University, Kyoto, Kyoto 606-8502, Japan}
\affiliation{Kavli Institute for the Physics and Mathematics of the Universe (WPI), The University of Tokyo Institutes for Advanced Study, University of Tokyo, Kashiwa, Chiba 277-8583, Japan}
\email{}

\begin{abstract}
Massive stars ($M>8\mathrm{M_\odot}$) emit neutrinos known as pre-supernova (pre-SN) neutrinos through thermal and nuclear interactions for cooling the stellar core during the final stage of stellar evolution. 
Real-time monitoring of their pre-SN neutrino interaction rate offers a crucial opportunity to issue an early warning to a core-collapse supernova. 
Some neutrino detectors, including KamLAND and Super-Kamiokande already operate pre-SN alarm systems based on a statistically significant excess of the observed event rate over the expected background.
To improve alarm sensitivity, an alarm method which incorporates the time evolution of the observed pre-SN neutrino event rate was proposed in \cite{ShapeAnalysis2021}.
We evaluate the performance of the light-curve likelihood approach under realistic KamLAND and SK operating conditions, including realistic background rates, global false-alarm-rate calibration, and the combined alarm.
The results demonstrate a significant improvement in the alarm time and distance compared to the conventional rate-only method, while maintaining the same false alarm rate.
\end{abstract}

\section{Introduction}

Core-collapse supernovae (CCSNe), in contrast to thermonuclear Type-Ia events, release a huge amount of energy in the form of neutrinos, gravitational waves and electromagnetic waves. 
The neutrino burst from a CCSN directly probes the stellar interior since neutrinos interact only via the weak force, whereas photons remain trapped in the envelope until much later.
In 1987, Kamiokande-II (\cite{Kamiokande-II_SN}), IMB (\cite{IMB_SN}) and Baksan (\cite{Baksan_SN}) achieved the first detection of the neutrino burst from SN1987A in the Large Magellanic Cloud, at a distance of $\sim 50\,\mathrm{kpc}$ (\cite{2019Natur.567..200P}).
The time and energy distributions of the 24 events observed by the three detectors provided a standard picture of the explosion mechanism (\cite{SN1987A_comb_analysis}).
Current and future neutrino detectors are expected to yield more detailed understanding of the explosion mechanism from the next galactic CCSN (\cite{Kate_SN}).

A massive star with initial mass $\geq 8\,\mathrm{M}_\odot$ at the Zero Age Main Sequence (ZAMS) emits neutrinos of all flavors during the late stages of stellar evolution.
These ``pre-supernova'' (pre-SN) neutrinos are produced by thermal and nuclear processes in the stellar core. 
Their detections could provide an understanding of late-stage stellar evolution and the neutrino mass ordering (\cite{Kato2020_review}).
Since these neutrinos are emitted before the core collapse, they can be used as an early warning to CCSNe.

In the sub-MeV to few-MeV energy range, relevant for pre-SN neutrino detection, the neutrino interactions are predominantly coherent neutrino-nucleus scattering, electron scattering and inverse beta decay.
Inverse beta decay (IBD, $\bar{\nu}_e+p\rightarrow e^++n$) which has an energy threshold of $1.806\,\mathrm{MeV}$, allows pre-SN electron antineutrinos to be observed by some large-volume neutrino detectors including  KamLAND (\cite{KamLAND_detector}), Super-Kamiokande (\cite{Super-Kamiokande:2002weg}) loaded with gadolinium (SK-Gd, \cite{lucaspaper}), JUNO (\cite{JUNO_preSN}) and SNO+ (\cite{SNO+}), which have sufficient sensitivity to few-MeV signals.
Furthermore, upcoming detectors such as DUNE (\cite{DUNE_SN}), Hyper-Kamiokande (\cite{HyperK}) and $100\,\mathrm{ton}$-scale dark matter detectors (\cite{Raj:2019wpy}) are expected to be sensitive to pre-SN neutrinos. 
To unite these global efforts, the SuperNova Early Warning System (SNEWS 2.0), an upgraded global network of neutrino observatories, aims to coordinate and disseminate pre-SN and supernova alerts (\cite{SNEWS}, \cite{SNEWS_2.0}).

KamLAND (\cite{kamland_preSN}) and Super-Kamiokande (\cite{lucaspaper}) deployed pre-SN early-warning systems in 2015 and 2021, respectively. 
Furthermore, a combined pre-SN alarm system integrating KamLAND and Super-Kamiokande was launched in $2023$ (\cite{CombinedAlarm2024}).\footnote{The web page of combined pre-SN alarm system between KamLAND and SK: \url{https://www.lowbg.org/presnalarm/}}
These systems issue an alarm with the statistical significance of a rate excess over the expected background.
\cite{ShapeAnalysis2021} proposed an approach that incorporates time evolution of neutrino luminosity, and demonstrated improved sensitivity relative to rate-only approach. 
In this work, we re-evaluate the alarm performance to pre-SN neutrinos by applying actual event selection criteria, under the realistic background conditions.
Furthermore, the alarm sensitivity is evaluated based on the false alarm rate taking continuous online operation into account.

The structure of this paper is as follows.
Section \ref{sec:presupernova model} introduces the pre-SN neutrino models used in this study.
Sections \ref{sec:detector} and \ref{sec:background} describe the configurations and background conditions of KamLAND and Super-Kamiokande,  respectively. Section \ref{sec:conventional_alarm} reviews the conventional alarm methods.
In Section \ref{sec:alarmmethodology}, we review the light-curve-based approach.
The performance results are reported in Section \ref{sec:result} and discussed in Section \ref{sec:discussion}.
Finally, Section \ref{sec:conclusion} summarizes the conclusions.

\section{Pre-supernova neutrino light curve model}\label{sec:presupernova model}

The evolution of massive stars proceeds through successive nuclear burning stages of hydrogen, helium, carbon, neon and silicon.
After the carbon-burning stage, neutrinos dominate the energy transport from the core, where the temperatures reach over $\mathrm{\sim10^8\,\mathrm{K}}$.
While all flavors of neutrino and antineutrino pairs are produced thermally via $e^-+e^+\rightarrow\nu+\bar{\nu}$, weak nuclear interactions such as beta decays and electron/positron captures become dominant after the Si-burning stage.

For this analysis, the expected pre-SN neutrino rate in each detector is evaluated using established pre-SN neutrino light curve models including \cite{odrzywolek2}, \cite{odrzywolek}, \cite{Kato2015}, \cite{Yoshida2016}, \cite{kato_preSN} and \cite{patton}.
These models are based on numerical one dimensional hydrodynamical simulations of the late stages of stellar evolution and have advanced our understanding of the pre-SN neutrino flux and spectrum.

\begin{figure}
    \centering
    \includegraphics[width=\linewidth]{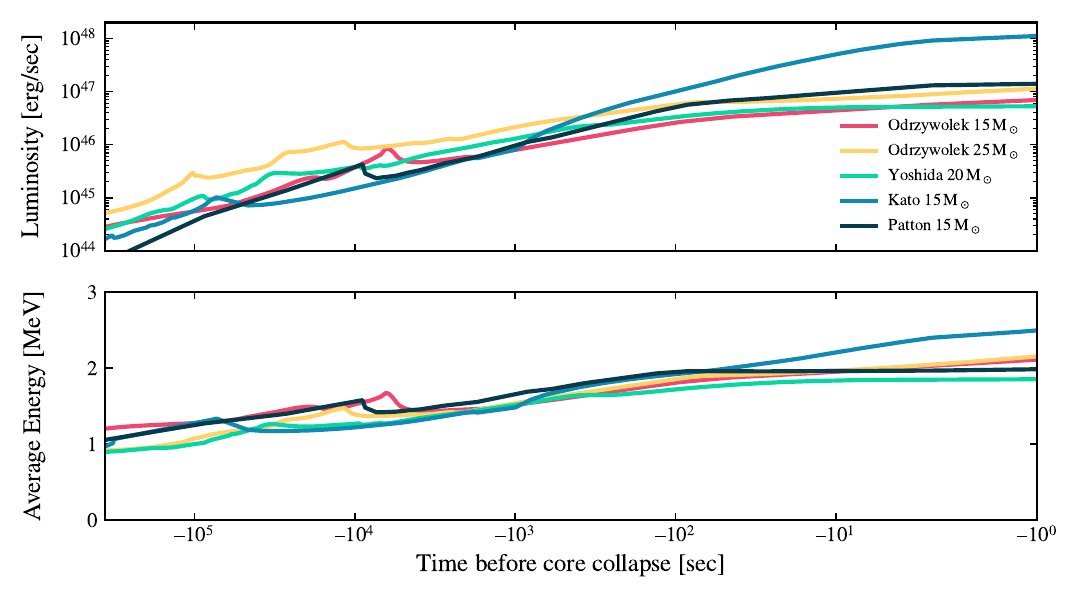}
    \caption{Electron antineutrino luminosity and their average energy are shown as a function of time before core collapse (horizontal axis). Oxygen- and Silicon-shell burning produce peaks around $-10^5$ s and $-10^4$ s before core collapse, respectively. The pre-SN neutrino light curve models are summarized in Table \ref{table:ReferenceModel}.}
    \label{fig:LuminosityEnergyAvg}
\end{figure}

The Odrzywolek model (\cite{odrzywolek2}, \cite{odrzywolek}) provides the neutrino flux which comes from thermal pair production. 
The production rate is based on the temperature, density and electron fraction under the nuclear statistical equilibrium as obtained from stellar-evolution calculations (\cite{Woosley_stellar}).
Since that implementation explicitly provides only the electron-(anti)neutrino component, the non-electron-(anti)neutrino flux is assumed to be $0.19$ times that of the electron-(anti)neutrinos.
The Yoshida (\cite{Yoshida2016}) and Kato models (\cite{kato_preSN}) use the same stellar evolution code calculated by \cite{Takahashi2013} and \cite{Takahashi2016}, but differ in their treatment of microphysics.
The Yoshida model considers only thermal pair production.
In contrast, the Kato model performs a more comprehensive calculation that includes nuclear weak interactions, such as beta decay and electron/positron captures for 3928 nuclei calculated via post-processing (\cite{Kato2020_review}), in addition to thermal processes.
Similarly, the Patton model (\cite{patton}) incorporates both thermal pair production and nuclear weak interactions.
This model evaluates the neutrino energy spectra using the stellar evolution code MESA (Modules for Experiments in Stellar Astrophysics), which couples the hydrodynamics with a 204-isotope nuclear network (\cite{MESA_reference}).
Table \ref{table:ReferenceModel} summarizes these pre-SN models.
Figure \ref{fig:LuminosityEnergyAvg} shows the corresponding simulated light curves and energy spectra. 
The exhaustion of nuclear fuel leads to core contraction, which ignites the surrounding shell.
At the onset of each new shell burning phase, core expansion temporarily reduces the central temperature and density (\cite{Kato2020_review}). 
This decrease leads to a suppression of the neutrino emission from the core.
The peaks and dips of the light curve and energy in Figure \ref{fig:LuminosityEnergyAvg} are associated with the oxygen- and the silicon-shell burning.

 Pre-SN $\bar{\nu}_e$ undergo flavor conversion via the Mikheyev-Smirnov-Wolfenstein (MSW) effect in the high-density environment of the stellar envelope. 
 This flavor conversion occurs because the effective mass eigenstates in matter differ from those in vacuum, and the resonance conditions in the MSW effect enhance the transition probabilities.
As a result, a portion of the original $\bar{\nu}_e$ flux converts into the other flavor neutrinos, and vice versa.
The observable $\bar{\nu}_e$ flux ($F_{\bar{\nu}_e}$) is $F_{\bar{\nu}_e}=pF^0_{\bar{\nu}_e}+(1-p)F^0_{\bar{\nu}_x}$, where $F^0_{\bar{\nu}_e}$ ($F^0_{\bar{\nu}_x}$) denotes the original (non)-electron antineutrino flux.
Given the mixing angles $\theta_{12}=33.76^\circ$ and $\theta_{13}=8.62^\circ$(\cite{NuFit2026}), the survival probability is $p=\cos^2\theta_{12}\cos^2\theta_{13}=0.68$ for the normal neutrino mass ordering (NO) and $p=\sin^2\theta_{13}=0.022$ for the inverted neutrino mass ordering (IO).
\floattable
\begin{deluxetable}{cccc}
    \tablecaption{Representative pre-SN light curve models \label{table:ReferenceModel}}
    \tablehead{\colhead{Model name} & \colhead{Stellar evolution code} & \colhead{Included reaction channels} \\}
    \startdata
        Odrzywolek (2010)& Woosley et al. (2002)            &Pair production\\
        Yoshida (2016)   & Takahashi et al. (2013 \& 2016)  &Pair production\\
        Kato (2017)      & Takahashi et al. (2013 \& 2016)  &Pair production+nuclear interaction\\
        Patton (2017)    & MESA                             &Pair production+nuclear interaction\\
    \enddata
\end{deluxetable}

\section{Detector}\label{sec:detector}

\subsection{KamLAND}
The Kamioka Liquid scintillator Anti-Neutrino Detector (KamLAND) is located $1,000\,\mathrm{m}$ beneath Mt. Ikenoyama ($36.42^\circ\mathrm{N}$, $137.31^\circ\mathrm{E}$) in Japan.
KamLAND comprises $1\,\mathrm{kt}$ of purified organic liquid scintillator (KamLS), contained within a $13$-m-diameter spherical balloon made of the $135$-$\mathrm{\mu m}$-thick nylon/EVOH film. 
Scintillation light is collected by the array of 1325 17-inch and 554 20-inch photomultiplier tubes (PMTs), facing the detector center.
The event vertex is reconstructed using hit-time and charge information from the 17-inch PMTs, while energies are reconstructed using the combined information from the 17- and 20-inch PMTs.

KamLAND detects pre-SN $\bar{\nu}_e$ via IBD interaction with a $1.806\,\mathrm{MeV}$ threshold.
The prompt scintillation light is produced by positron and its annihilation $\gamma$s.
The electron antineutrino energy $E_{\bar{\nu}_e}$ is determined by the prompt energy ($E_\mathrm{prompt}$) including positron kinetic and annihilation energies ($2\times511\,\mathrm{keV}$): $E_{\bar{\nu}_e}\simeq E_\mathrm{prompt}+\tilde{E}_n+0.8\,\mathrm{MeV}$, where $\tilde{E}_n$ is the average recoil energy ($\sim\mathcal{O}(10)\,\mathrm{keV}$).
The thermal-neutrons are captured with a mean lifetime of $207.5\pm2.8\,\mathrm{\mu s}$ on protons (emitting $2.2\,\mathrm{MeV}$ $\gamma$) or carbon-12 nuclei (emitting $4.9\,\mathrm{MeV}$ $\gamma$), with relative fractions of $\sim 99\%$ and $\sim 1\%$, respectively.
The sub-MeV energy threshold of KamLAND is essential to explore delayed coincidence signatures, between prompt positron signal and delayed thermal-neutron capture signals, thereby achieving highly effective background rejection.

For the neutrinoless double-beta decay search (KamLAND-Zen), a mini-balloon containing xenon-loaded liquid scintillator was installed in two phases: a 3.08-m-diameter balloon (2011--2015; KamLAND-Zen 400; \cite{KamLANDZen}) and a 3.80-m-diameter balloon (2019--2024; KamLAND-Zen 800; \cite{KamLANDZen2025}).
In the next phase, a new mini-balloon  is planned for KamLAND2-Zen (\cite{KamLANDZen2025}). 
We assume the mini-balloon will remain in place in future operation and therefore impose a mini-balloon cut on delayed events to remove backgrounds from the balloon and its support materials. 

IBD candidates are selected using spatial correlation ($\Delta R$), time difference ($\Delta T$) between prompt and delayed signals, prompt energy ($E_\mathrm{prompt}$) and delayed energy ($E_\mathrm{delayed}$). 
The criteria are $0.9\,\mathrm{MeV}<E_\mathrm{prompt}<4.0\,\mathrm{MeV}$, $1.8\,\mathrm{MeV}<E_\mathrm{delayed}<2.6\,\mathrm{MeV}$ or $4.4\,\mathrm{MeV}<E_\mathrm{delayed}<5.6\,\mathrm{MeV}$, $\Delta R<200\,\mathrm{cm}$ and $0.5\,\mathrm{\mu s}<\Delta T<1,000\,\mathrm{\mu s}$.
A fiducial volume cut is applied to exclude regions $>600\,\mathrm{cm}$ from the center of the detector in order to remove the accidental coincidences that increase near the outer balloon surface ($R=650\,\mathrm{cm}$).
The delayed events, reconstructed within the mini-balloon, are also cut. 
Additionally, a likelihood cut is applied to suppress accidental coincidences.

\subsection{Super-Kamiokande}
Super-Kamiokande (SK) is a water-Cherenkov detector, located, like KamLAND, $\sim1,000\,\mathrm{m}$ underground beneath Mt. Ikenoyama.
It consists of a cylindrical stainless steel tank, $39.3\,\mathrm{m}$ in diameter and $41.4\,\mathrm{m}$ in height, originally filled with $50\,\mathrm{kt}$ of water (\cite{Super-Kamiokande:2002weg}).
The main detector is an inner cylinder $33.8\,\mathrm{m}$ in diameter and $36.2 \,\mathrm{m}$ in height, containing approximately $32\,\mathrm{kt}$ of water. 
The standard fiducial volume of $22.5\,\mathrm{kt}$ is defined as the region more than $2.0\,\mathrm{m}$ away from the walls of the inner detector.
Approximately 11,000 inward-facing 20-inch PMTs detect water Cherenkov light from neutrino interactions in the water.

In July 2020, SK began a new phase, SK-Gd, after dissolving gadolinium sulfate octahydrate ($\mathrm{Gd}_2(\mathrm{SO}_4)_3\cdot8\mathrm{H}_2\mathrm{O}$) to the water in the detector.
Gadolinium has an exceptionally large thermal-neutron capture cross section ($\sim49,000\,\mathrm{barns}$, compared to $\sim0.3\,\mathrm{barns}$ for hydrogen). Neutron capture on gadolinium produces a more energetic $\sim8\,\mathrm{MeV}$ $\gamma$ cascade, whereas capture on hydrogen emits a single $2.2\,\mathrm{MeV}$ $\gamma$, enabling efficient neutron identification.
Two loading campaigns increased the $\mathrm{Gd}$ concentration to $0.01$\% (July--August 2020; \cite{Super-Kamiokande:2021the}) and to $0.03$\% (May--July 2022; \cite{2024NIMPA106569480A}), enhancing the thermal-neutron capture efficiency to $\simeq75$\%.

With enhanced thermal-neutron capture rate on $\mathrm{Gd}$, SK can search for pre-SN $\bar{\nu}_e$ via IBD using a delayed coincidence between the prompt positron and delayed neutron capture, requiring $\Delta R<300\,\mathrm{cm}$ and $\Delta T<80\,\mathrm{\mu s}$.
IBD event selection is based on two different levels of boosted decision tree (BDT): the pre-selection $\mathrm{BDT_{online}}$ and the final selection $\mathrm{BDT_{offline}}$, balancing between the computational speed and the classification performance (\cite{lucaspaper}).

\section{Background}\label{sec:background}

Pre-SN neutrino searches face several background sources: electron antineutrinos from nuclear reactors, electron antineutrinos from radioactive decays within the Earth (geoneutrinos), and non-neutrino processes such as $(\alpha,n)$ reactions and accidental coincidences.

Reactor $\bar{\nu}_e$ events, which are one of the dominant backgrounds in this study, are produced from the beta decay of the fission products of neutron-rich $^{235} \mathrm{U}$, $^{238}\mathrm{U}$, $^{239}\mathrm{Pu}$ and $^{241}\mathrm{Pu}$ nuclides in reactors operating in Japan and Korea.
We calculate reactor $\bar{\nu}_e$ event rate based on theoretical spectrum (\cite{Huber2011}, \cite{Mueller2011} and \cite{Vogel1981}).
Although all Japanese reactors were shut down after the 2011 Great East Japan Earthquake under strengthened safety regulations, several units have gradually resumed operation since 2015.
We calculate reactor $\bar{\nu}_e$ event rate, by weighting each core with its reported real-time electric (thermal) power output.
The $\bar{\nu}_e$ survival probability is computed with these parameters: $\Delta m_{21}^2=7.53\times10^{-5}\,\mathrm{eV}^2$, $\tan^2\theta_{12}=0.436$ and $\sin^2\theta_{13}=0.023$ (\cite{KamLAND:2013rgu}).
Geoneutrino events, produced in decay chains of $^{238}\mathrm{U}$ and $^{232}\mathrm{Th}$ nuclei inside Earth, also contribute as an additional background. 
The expected geoneutrino event rate is calculated using the parameters from \cite{Enomoto_geo}.

$(\alpha,n)$ reactions originate from $\alpha$ emitters in detector materials and can yield two correlated signals that satisfy the IBD selection.
In particular, $^{13}\mathrm{C}(\alpha,n)^{16}\mathrm{O}$ in KamLAND (\cite{KamLAND_alphan}), $^{18}\mathrm{O}(\alpha,n)^{21}\mathrm{Ne}^*$ and $^{17}\mathrm{O}(\alpha,n)^{20}\mathrm{Ne}^*$ in SK constitute non-negligible backgrounds.
Although accidental coincidences are strongly suppressed, the residual events constitute a part of the main backgrounds.

Background rates for KamLAND and SK-Gd are taken from \cite{CombinedAlarm2024} under the “medium reactor-activity” scenario representative of 2024 operations, which assumes full-power running of nearby Japanese reactors (Mihama-3; Ohi-3,4; Takahama-1--4) and relevant Korean reactors.
The total background rates are $0.19\,\mathrm{day}^{-1}$ in KamLAND and $12.4\,\mathrm{day}^{-1}$ in SK.

\section{Conventional alarm method}\label{sec:conventional_alarm}

\begin{figure}[t]
    \centering
    \gridline{\fig{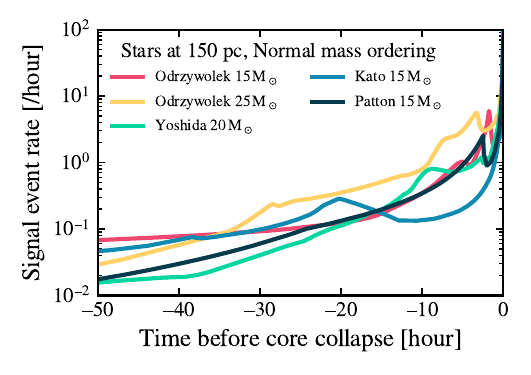}        {0.48\textwidth}{(a) KamLAND}
              \fig{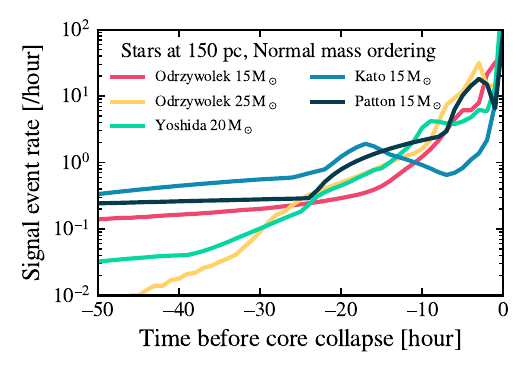}{0.48\textwidth}{(b) Super-Kamiokande}}
    \caption{\label{fig:Eventrate} Time evolution of pre-SN event rate in (a) KamLAND and (b) Super-Kamiokande for the stars at $150\,\mathrm{pc}$. The mass orderings are assumed to be normal. The pre-SN neutrino light curve models are summarized in Table \ref{table:ReferenceModel}.}
\end{figure}

As described in the previous section, KamLAND and SK detect pre-SN $\bar{\nu}_e$ via IBD and the expected IBD event rates are calculated from the $\bar{\nu}_e$ luminosity and energy spectrum.
The resulting time evolutions for both detectors are illustrated in Figure \ref{fig:Eventrate}.
Both KamLAND and SK early warning systems, for pre-SN neutrino searches, issue alarms based on the statistical significance of the excess in the observed event counts, evaluated in sliding time windows, relative to the expected background (rate analysis).
The analysis time window is optimized to $24\,\mathrm{h}$ for KamLAND and $12\,\mathrm{h}$ for SK.
The alarm performance characterizations are reported in (\cite{kamland_preSN}, \cite{lucaspaper} and \cite{CombinedAlarm2024}).

KamLAND enables early warning due to its lower background, whereas SK's larger target mass allows for a rapid increase in significance and sensitivity to distant stars. By leveraging these complementary advantages and compensating for each other's dead time, the combined alarm system achieves earlier alert times than individual detectors.
It likewise employs a rate-based analysis of the event counts in the two detectors and has operated since 2023 (\cite{CombinedAlarm2024}).
Combined alarm system issues an alarm earlier than individual detectors (\cite{CombinedAlarm2024}).

\section{Alarm methodology}\label{sec:alarmmethodology}

The alarm method using the neutrino light curve (rate+time analysis) formulated as a one-sided hypothesis test between background-only $\mathrm{H}_0$ and background+signal $\mathrm{H}_1$ (\cite{ShapeAnalysis2021}).
The resulting p-value quantifies the significance against $\mathrm{H}_0$. 
In the rate+time analysis, let the data within a sliding analysis window $[t_0-T,t_0]$ be the event count $n$ and the set of timestamps $\{t_i\}|_{i=1,2,\cdots ,n}$, where $t_0$ denotes the observation time (the end of the analysis window).

The expected pre-SN signal rate is a function of time relative to the core collapse, $R_\mathrm{s}(t-t*)$, where $t*$ is the start time of core collapse. 
The background rate is $R_B(t)$.
The expected signal event count $S(t*)$ and background count $B$ over the time window $T$ (from $t_0-T$) are the integrals of these rates:
\begin{equation}
    S^j(t*) = \int_{t_0-T}^{t_0} R_\mathrm{S}^j(t-t*) \,dt 
    \qquad \text{and} \qquad 
    B = \int_{t_0-T}^{t_0} R_\mathrm{B}(t) \,dt.
\end{equation}
where $j$ represents the pre-SN neutrino model.
Under the inhomogeneous-Poisson model, the likelihood functions factorize into a Poisson term for $n$ (\cite{ShapeAnalysis2021}):
\begin{equation}
    \mathcal{L}_\mathrm{BG+Signal}^j(n,\{t_i\},t_*) = \frac{(S^j(t_*)+B)^n}{n!}e^{-[S^j(t_*)+B]}\times\prod_{i=1}^n\frac{1}{S^j(t_*)+B}[R_\mathrm{S}^j(t_i-t*)+R_B(t_i)],
\end{equation}
\begin{equation}
    \mathcal{L}_\mathrm{BG\,Only}(n) = \frac{B^n}{n!}e^{-B}\times\prod_{i=1}^n\frac{R_B(t_i)}{B}.
\end{equation}
In order to discriminate background-only and background+signal hypothesis efficiently,  we use log likelihood ratio function ($\mathrm{LLR}$),
\begin{equation}
    \mathrm{LLR}^j(n,\{t_i\},t_*) = \mathrm{log} \frac{\mathcal{L}_\mathrm{BG+Signal}^j(n,\{t_i\},t_*)}{\mathcal{L}_\mathrm{BG\,Only}(n,\{t_i\})}.
\end{equation}
The $\mathrm{LLR}$ is profiled over the nuisance parameter $t_*$ to define the test statistic:
\begin{equation}\label{eq:test_statistic_R+T}
    \Lambda^j(n,\{t_i\}) = \max_{t_*}\,\mathrm{LLR}^j(n,\{t_i\},t_*).
\end{equation}

The online alarm system continuously calculates $\Lambda^j$ in sliding window.
The alarm system introduces a trials factor (look-elsewhere effect) that would otherwise inflate naive significances.
We therefore calibrate the global p-value and the false alarm rate (FAR) using toy Monte Carlo under $\mathrm{H}_0$ (background-only), measuring how often $\Lambda$ exceeds a threshold in continuous operation.
This statistical test depends on a reference pre-SN neutrino light-curve model, which represents $j$. 
In this study, we evaluate an ensemble of models and report the minimum  FAR across the model set:
\begin{equation}\label{eq:MinimizeFAR}
    \mathrm{FAR}=\min_{j}\,\mathrm{FAR}^j.
\end{equation}

For the combined alarm, the joint log-likelihood ratio ($\Lambda_\mathrm{combined}$) is given by the sum of the individual log-likelihood ratios, assuming independent data streams, and is profiled over a common $t_*$.
The FAR for the combined alarm is evaluated by applying the same toy Monte Carlo procedure to $\Lambda_\mathrm{combined}$.

\section{Alarm sensitivity to pre-supernova neutrinos}\label{sec:result}

The alarm performance using rate+time analysis is evaluated for KamLAND, SK and their combined sensitivity.
The reference model set used in the likelihood includes available models in the $15$--$25\,\mathrm{M_\odot}$ range.
The benchmark signal injections shown below, however, are restricted to $15\,\mathrm{M_\odot}$ models, following the setup of \cite{CombinedAlarm2024}.
These results should therefore be interpreted as representative performance benchmarks rather than a systematic survey over progenitor mass.

Signal injections are generated from the models by Odrzywolek and Patton, assuming a $15\,\mathrm{M}_\odot$ star at $150\,\mathrm{pc}$ with either normal or inverted neutrino mass ordering. 
The reference light-curve models used in the likelihood are those listed in Table \ref{table:ReferenceModel}.
We focus on the alarm performance for $\alpha$ Orionis (Betelgeuse), a nearby candidate for a future core collapse event.
Observational studies suggest that Betelgeuse has an initial mass in the range $17$--$25\,\mathrm{M}_\odot$ (\cite{Dolan2016}, \cite{Joyce_2020}).
Conservatively, we adopt $15$--$25\,\mathrm{M_\odot}$ for the reference.
Due to our incomplete understanding of stellar evolution, uncertainties still remain in modeling true pre-SN signals. Therefore, we perform a conservative sensitivity evaluation considering the degradation of efficiency when the injected true pre-SN  light curve model differs from the reference models.
Unless otherwise noted, the reference assumes a star at a fixed distance of $150\,\mathrm{pc}$ with normal neutrino mass ordering, because it predicts a higher survival probability than inverted neutrino mass ordering, resulting in a larger observable event rate while retaining similar temporal features.

\begin{figure}[t]
    \centering
    \gridline{\fig{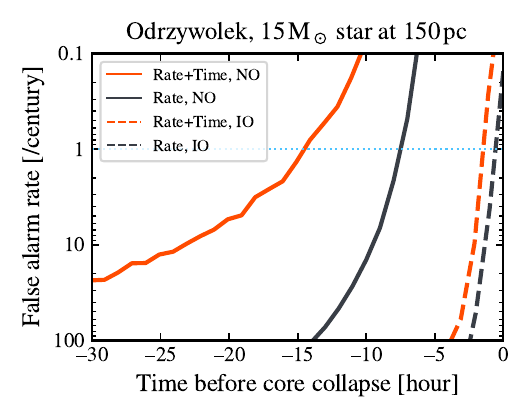}{0.48\textwidth}{(a) Odrzywolek model signal}
              \fig{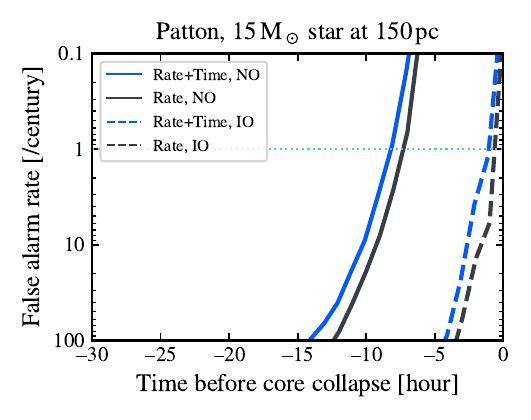}    {0.48\textwidth}{(b) Patton model signal}}
    \caption{\label{fig:KLAlarmPerformance} Time evolution of expected FAR in KamLAND. (a) and (b) correspond to the Patton and Odrzywolek models, respectively, used as the signal injection models for a $15\,\mathrm{M}_\odot$ star at $150\,\mathrm{pc}$. Reference pre-SN light-curve models correspond to $15$--$25\,\mathrm{M}_\odot$ stars at $150\,\mathrm{pc}$ assuming the normal mass ordering, as listed in Table \ref{table:ReferenceModel}. The horizontal axis represents the minimized FAR for these reference models, as shown in \autoref{eq:MinimizeFAR}. The horizontal dashed light-blue line represents FAR of $1\,\mathrm{/century}$.}
\end{figure}

\begin{figure}[t]
    \centering
    \gridline{\fig{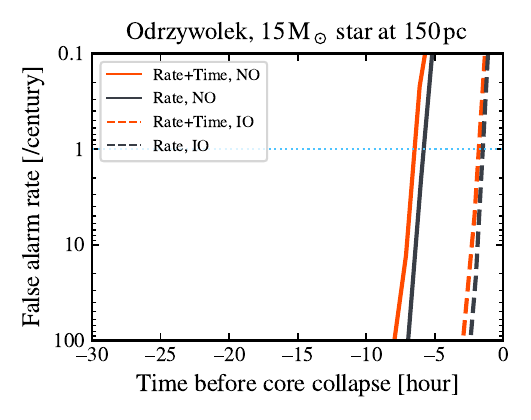}{0.48\textwidth}{(a) Odrzywolek model signal}
              \fig{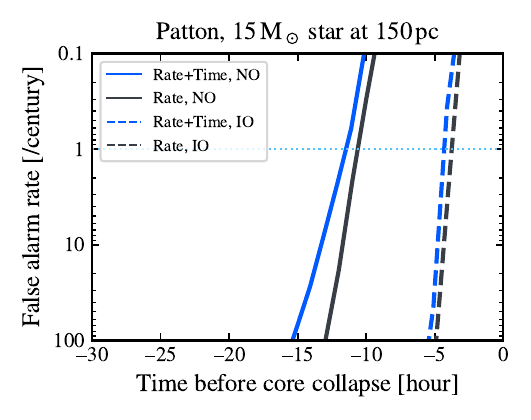}    {0.48\textwidth}{(b) Patton model signal}}
    \caption{\label{fig:SKAlarmPerformance} Time evolution of expected FAR in SK. The pre-SN neutrino models and lines used in these plots are described in the caption of Figure \ref{fig:KLAlarmPerformance}.}
\end{figure}

Figure \ref{fig:KLAlarmPerformance} compares the time evolution of the expected FAR between rate-only and rate+time analysis for KamLAND.
The analysis time window for the rate+time analysis is set to $200\,\mathrm{h}$, roughly spanning from the onset of silicon burning to core collapse.
The horizontal dashed line marks a FAR of one per century.
For an injection based on the Odrzywolek model with normal neutrino mass ordering, the KamLAND rate+time method achieves the best performance among the tested configurations and issues an alert $\geq 14.5\,\mathrm{h}$ before core collapse.
Figure \ref{fig:SKAlarmPerformance} shows the SK FAR comparison. 
The analysis time window for the SK rate+time analysis is set to $48\,\mathrm{h}$.
In the most favorable case, the SK rate+time analysis issues an alert $\geq 11.6\,\mathrm{h}$ before core collapse.

\begin{figure}[t]
    \centering
    \gridline{\fig{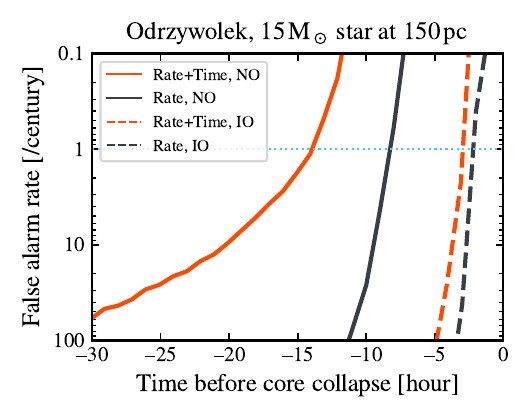}{0.48\textwidth}{(a) Odrzywolek model signal}
              \fig{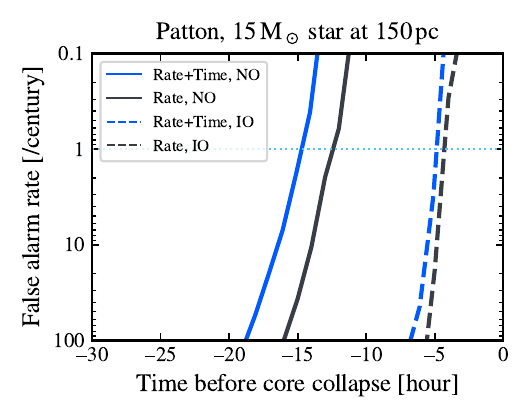}    {0.48\textwidth}{(b) Patton model signal}}
    \caption{\label{fig:CombinedAlarmPerformance}Time evolution of expected FAR in combined alarm case. The pre-SN neutrino models and lines used in these plots are described in the caption of Figure \ref{fig:KLAlarmPerformance}.}
\end{figure}

For the combined KamLAND and SK alarm, sensitivities are evaluated with independent time-series Monte Carlo for each detector. 
The resulting FAR is shown in Figure \ref{fig:CombinedAlarmPerformance}.
The combined rate+time analysis improves sensitivity and delivers earlier alerts than either detector alone.  

\begin{figure}[t]
    \centering
    \gridline{\fig{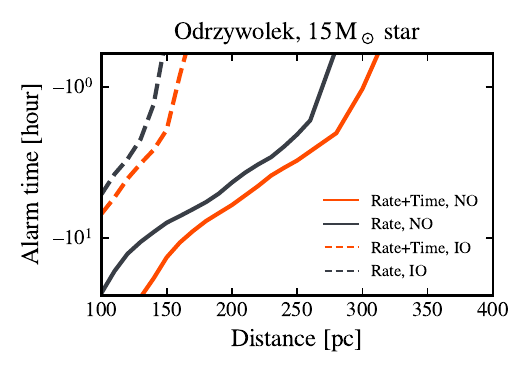}{0.48\textwidth}{(a) KamLAND, Odrzywolek model signal}
              \fig{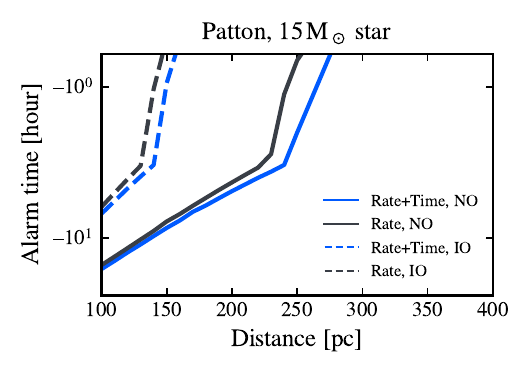}    {0.48\textwidth}{(b) KamLAND, Patton model signal}}
    \gridline{\fig{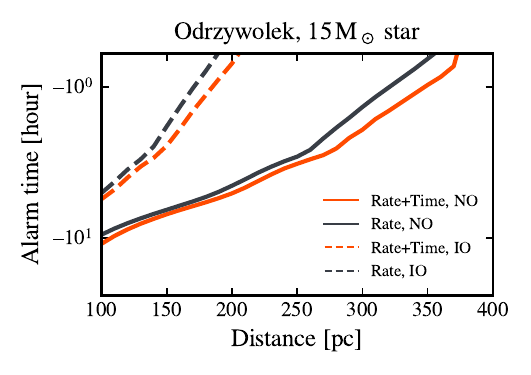}{0.48\textwidth}{(c) SK, Odrzywolek model signal}
              \fig{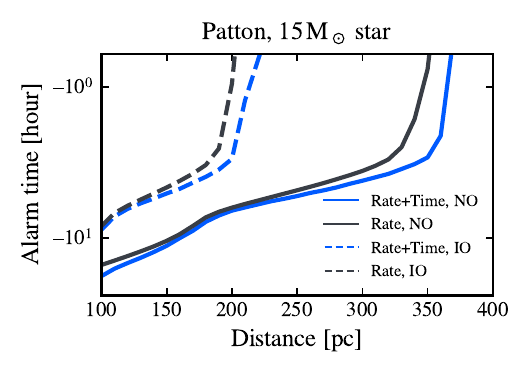}    {0.48\textwidth}{(d) SK, Patton model signal}}
    \gridline{\fig{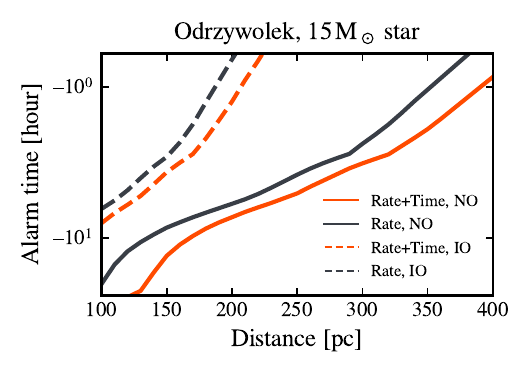}{0.48\textwidth}{(e) Combined, Odrzywolek model signal}
              \fig{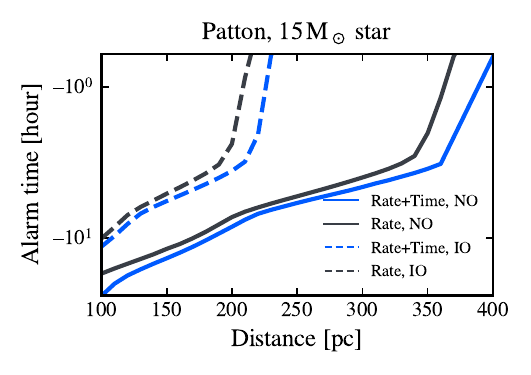}    {0.48\textwidth}{(f) Combined, Patton model signal}}
    \caption{\label{fig:AlarmDistance}The distance dependence on warning time for a $15\,\mathrm{M_\odot}$ star.
    Reference pre-SN light-curve models correspond to $15$--$25\,\mathrm{M}_\odot$ stars at $150\,\mathrm{pc}$ assuming the normal mass ordering, as listed in Table \ref{table:ReferenceModel}.
    The alarm threshold is FAR of $1\,\mathrm{/century}$.}
\end{figure}

Figure \ref{fig:AlarmDistance} shows the distance dependence of warning time for $15\,\mathrm{M_\odot}$ stars for KamLAND, SK and their combined alarm case.
Rate+time analysis can issue alarm for more distant CCSNe.

\section{Discussion}\label{sec:discussion}

Table \ref{table:AlarmTime} summarizes the alarm lead times (hours before core collapse) for KamLAND, SK, and their combined configuration.
For each individual detector, the rate+time analysis achieves lead times comparable to those of the combined rate-only analysis.
Even if one detector is inoperative, the other maintains performance comparable to the conventional combined (rate-only) system.
Furthermore, the combined rate+time analysis yields the earliest alerts among all configurations:
For a $15\,\mathrm{M_\odot}$ star at $150\,\mathrm{pc}$, assuming normal neutrino mass ordering, it provides a warning $14.0\,\mathrm{h}$ (Odrzywolek model) and $14.7\,\mathrm{h}$ (Patton model) before core collapse whereas rate analysis yields $8.2\,\mathrm{h}$ and $12.3\,\mathrm{h}$ respectively.

Beyond earlier alerts, the rate+time analysis enhances sensitivity to more distant sources relative to rate-only methods.
Since the pre-SN neutrino signal decreases with the square of the distance, leveraging temporal information allows statistically significant alerts with fewer events, enabling detection of weaker signals that may be inaccessible to rate-only analyses.

\begin{table*}[htbp]
    \centering
    \caption{Warning times [hours] before core collapse of $15\,\mathrm{M}_\odot$ stars at $150\,\mathrm{pc}$. Reference pre-SN light-curve models correspond to $15$--$25\,\mathrm{M}_\odot$ stars at $150\,\mathrm{pc}$ assuming the normal mass ordering, as listed in Table \ref{table:ReferenceModel}. The alarm threshold is a FAR of $1\,\mathrm{century}^{-1}$.
    \label{table:AlarmTime}}
    \begin{tabular*}{0.9\textwidth}{@{\extracolsep{\fill}}ccccccc}
        \hline\hline
        Detector & Analysis & \multicolumn{4}{c}{Warning time [hour]}  \\
         & & \multicolumn{2}{c}{Odrzywolek $15\,\mathrm{M}_\odot$} & \multicolumn{2}{c}{Patton $15\,\mathrm{M}_\odot$}  \\
         & & NO & IO & NO & IO\\
        \hline
        KamLAND & Rate      & 7.5  & N/A & 7.3  & 0.2\\
                & Rate+time & 14.5 & 1.6 & 8.2  & 0.9\\
        SK      & Rate      & 5.8  & 1.6 & 10.5 & 3.8\\
                & Rate+time & 6.3  & 1.9 & 11.6 & 4.3\\
        Combined& Rate      & 8.2  & 2.1 & 12.3 & 4.3\\
                & Rate+time & 14.0 & 2.8 & 14.7 & 4.8\\
        \hline
    \end{tabular*}
\end{table*}

\section{Summary}\label{sec:conclusion}

We evaluate the alarm sensitivity using the time evolution of the pre-SN neutrino event rate (``rate+time'' analysis) under realistic conditions of detector operation and data analysis in KamLAND and SK.
To represent realistic background conditions, background rates are calculated assuming full-power operation of Japanese commercial reactors near the Kamioka mine (Mihama-3; Ohi-3,4; Takahama-1--4) and relevant Korean reactors.
The alarm method uses multiple time profiles, derived from representative pre-SN neutrino light curve models \cite{odrzywolek2}, \cite{odrzywolek}, \cite{Yoshida2016}, \cite{kato_preSN} and \cite{patton} in the mass range of $15$--$25\,\mathrm{M_\odot}$, to ensure robustness. 

We also evaluated the alarm sensitivity of rate+time analysis for a $15\,\mathrm{M_\odot}$ star at $150\,\mathrm{pc}$ in KamLAND, SK and their combined configuration.
The analysis time window of rate+time analysis is $200\,\mathrm{h}$ in KamLAND and $48\,\mathrm{h}$ in SK.
The rate+time analysis issues earlier alarms than the conventional rate-only method.
The optimistic alarm time is $14.5\,\mathrm{h}$ and $11.6\,\mathrm{h}$ before core collapse in KamLAND and SK, respectively.
Using the combined system with the rate+time analysis yields the earliest alerts overall, up to $14.7\,\mathrm{h}$ in the optimistic case, compared to $12.3\,\mathrm{h}$ for the combined rate-only analysis.

\begin{acknowledgements}
This work is supported by JSPS KAKENHI Grant Numbers 24H02237, 24H02242, 25KJ0557, and 24H00243, the Science and Technology Facilities Council (STFC).
We would like to thank all KamLAND and Super-Kamiokande collaborators for supporting this effort.
\end{acknowledgements}

\bibliography{main}

@article{Kamiokande-II_SN,
    author = "Hirata, K. and others",
    editor = "Wali, K. C.",
    collaboration = "Kamiokande-II",
    title = "{Observation of a Neutrino Burst from the Supernova SN 1987a}",
    reportNumber = "UT-ICEPP-87-01, UPR-142E",
    doi = "10.1103/PhysRevLett.58.1490",
    journal = "Phys. Rev. Lett.",
    volume = "58",
    pages = "1490--1493",
    year = "1987"
}

@article{IMB_SN,
    author = "Bionta, R. M. and others",
    title = "{Observation of a Neutrino Burst in Coincidence with Supernova SN 1987a in the Large Magellanic Cloud}",
    reportNumber = "UCI-NEUTRINO-87-10",
    doi = "10.1103/PhysRevLett.58.1494",
    journal = "Phys. Rev. Lett.",
    volume = "58",
    pages = "1494",
    year = "1987"
}

@article{Baksan_SN,
    author = "Alekseev, E. N. and Alekseeva, L. N. and Volchenko, V. I. and Krivosheina, I. V.",
    editor = "Tran Thanh Van, J.",
    title = "{Possible Detection of a Neutrino Signal on 23 February 1987 at the Baksan Underground Scintillation Telescope of the Institute of Nuclear Research}",
    journal = "JETP Lett.",
    volume = "45",
    pages = "589--592",
    year = "1987"
}

@ARTICLE{SN1987A_comb_analysis,
       author = {{Pagliaroli}, G. and {Vissani}, F. and {Costantini}, M.~L. and {Ianni}, A.},
        title = "{Improved analysis of SN1987A antineutrino events}",
      journal = {Astroparticle Physics},
         year = 2009,
        month = apr,
       volume = {31},
       number = {3},
        pages = {163-176},
          doi = {10.1016/j.astropartphys.2008.12.010},
archivePrefix = {arXiv},
       eprint = {0810.0466},
 primaryClass = {astro-ph},
       adsurl = {https://ui.adsabs.harvard.edu/abs/2009APh....31..163P}
}

@ARTICLE{Kate_SN,
       author = {{Scholberg}, Kate},
        title = "{Supernova Neutrino Detection}",
      journal = {Annual Review of Nuclear and Particle Science},
         year = 2012,
        month = nov,
       volume = {62},
        pages = {81-103},
          doi = {10.1146/annurev-nucl-102711-095006},
archivePrefix = {arXiv},
       eprint = {1205.6003},
 primaryClass = {astro-ph.IM},
       adsurl = {https://ui.adsabs.harvard.edu/abs/2012ARNPS..62...81S}
}

@ARTICLE{2019Natur.567..200P,
       author = {{Pietrzy{\'n}ski}, G. and {Graczyk}, D. and {Gallenne}, A. and {Gieren}, W. and {Thompson}, I.~B. and {Pilecki}, B. and {Karczmarek}, P. and {G{\'o}rski}, M. and {Suchomska}, K. and {Taormina}, M. and {Zgirski}, B. and {Wielg{\'o}rski}, P. and {Ko{\l}aczkowski}, Z. and {Konorski}, P. and {Villanova}, S. and {Nardetto}, N. and {Kervella}, P. and {Bresolin}, F. and {Kudritzki}, R.~P. and {Storm}, J. and {Smolec}, R. and {Narloch}, W.},
        title = "{A distance to the Large Magellanic Cloud that is precise to one per cent}",
      journal = {\nat},
         year = 2019,
        month = mar,
       volume = {567},
       number = {7747},
        pages = {200-203},
          doi = {10.1038/s41586-019-0999-4},
archivePrefix = {arXiv},
       eprint = {1903.08096},
 primaryClass = {astro-ph.GA},
       adsurl = {https://ui.adsabs.harvard.edu/abs/2019Natur.567..200P}
}

@article{Raj:2019wpy,
    author = "Raj, Nirmal and Takhistov, Volodymyr and Witte, Samuel J.",
    title = "{Presupernova neutrinos in large dark matter direct detection experiments}",
    eprint = "1905.09283",
    archivePrefix = "arXiv",
    primaryClass = "hep-ph",
    doi = "10.1103/PhysRevD.101.043008",
    journal = "Phys. Rev. D",
    volume = "101",
    number = "4",
    pages = "043008",
    year = "2020"
}

@article{DUNE_SN,
    author = "Abi, B. and others",
    collaboration = "DUNE",
    title = "{Supernova neutrino burst detection with the Deep Underground Neutrino Experiment}",
    eprint = "2008.06647",
    archivePrefix = "arXiv",
    primaryClass = "hep-ex",
    reportNumber = "FERMILAB-PUB-20-380-LBNF, FERMILAB-PUB-20-380-LBNF",
    doi = "10.1140/epjc/s10052-021-09166-w",
    journal = "Eur. Phys. J. C",
    volume = "81",
    number = "5",
    pages = "423",
    year = "2021"
}

@article{Woosley_stellar,
    author = "Woosley, S. E. and Heger, A. and Weaver, T. A.",
    title = "{The evolution and explosion of massive stars}",
    doi = "10.1103/RevModPhys.74.1015",
    journal = "Rev. Mod. Phys.",
    volume = "74",
    pages = "1015--1071",
    year = "2002"
}

@ARTICLE{Takahashi2013,
       author = {{Takahashi}, Koh and {Yoshida}, Takashi and {Umeda}, Hideyuki},
        title = "{Evolution of Progenitors for Electron Capture Supernovae}",
      journal = {\apj},
         year = 2013,
        month = jul,
       volume = {771},
       number = {1},
          eid = {28},
        pages = {28},
          doi = {10.1088/0004-637X/771/1/28},
archivePrefix = {arXiv},
       eprint = {1302.6402},
 primaryClass = {astro-ph.SR},
       adsurl = {https://ui.adsabs.harvard.edu/abs/2013ApJ...771...28T}
}

@ARTICLE{Takahashi2016,
       author = {{Takahashi}, Koh and {Yoshida}, Takashi and {Umeda}, Hideyuki and {Sumiyoshi}, Kohsuke and {Yamada}, Shoichi},
        title = "{Exact and approximate expressions of energy generation rates and their impact on the explosion properties of pair instability supernovae}",
      journal = {\mnras},
         year = 2016,
        month = feb,
       volume = {456},
       number = {2},
        pages = {1320-1331},
          doi = {10.1093/mnras/stv2649},
archivePrefix = {arXiv},
       eprint = {1511.03040},
 primaryClass = {astro-ph.SR},
       adsurl = {https://ui.adsabs.harvard.edu/abs/2016MNRAS.456.1320T}
}

@ARTICLE{Dolan2016,
       author = {{Dolan}, Michelle M. and {Mathews}, Grant J. and {Lam}, Doan Duc and {Quynh Lan}, Nguyen and {Herczeg}, Gregory J. and {Dearborn}, David S.~P.},
        title = "{Evolutionary Tracks for Betelgeuse}",
      journal = {\apj},
         year = 2016,
        month = mar,
       volume = {819},
       number = {1},
          eid = {7},
        pages = {7},
          doi = {10.3847/0004-637X/819/1/7},
archivePrefix = {arXiv},
       eprint = {1406.3143},
 primaryClass = {astro-ph.SR},
       adsurl = {https://ui.adsabs.harvard.edu/abs/2016ApJ...819....7D}
}

@article{Joyce_2020,
doi = {10.3847/1538-4357/abb8db},
url = {https://dx.doi.org/10.3847/1538-4357/abb8db},
year = {2020},
month = {oct},
publisher = {The American Astronomical Society},
volume = {902},
number = {1},
pages = {63},
author = {Meridith Joyce and Shing-Chi Leung and László Molnár and Michael Ireland and Chiaki Kobayashi and Ken’ichi Nomoto},
title = {Standing on the Shoulders of Giants: New Mass and Distance Estimates for Betelgeuse through Combined Evolutionary, Asteroseismic, and Hydrodynamic Simulations with MESA},
journal = {The Astrophysical Journal}
}

@ARTICLE{SNO+,
       author = {{Albanese}, V. and {Alves}, R. and {Anderson}, M.~R. and {Andringa}, S. and {Anselmo}, L. and {Arushanova}, E. and {Asahi}, S. and {Askins}, M. and {Auty}, D.~J. and {Back}, A.~R. and {Back}, S. and {Bar{\~a}o}, F. and {Barnard}, Z. and {Barr}, A. and {Barros}, N. and {Bartlett}, D. and {Bayes}, R. and {Beaudoin}, C. and {Beier}, E.~W. and {Berardi}, G. and {Bialek}, A. and {Biller}, S.~D. and {Blucher}, E. and {Bonventre}, R. and {Boulay}, M. and {Braid}, D. and {Caden}, E. and {Callaghan}, E.~J. and {Caravaca}, J. and {Carvalho}, J. and {Cavalli}, L. and {Chauhan}, D. and {Chen}, M. and {Chkvorets}, O. and {Clark}, K.~J. and {Cleveland}, B. and {Connors}, C. and {Cookman}, D. and {Coulter}, I.~T. and {Cox}, M.~A. and {Cressy}, D. and {Dai}, X. and {Darrach}, C. and {Davis-Purcell}, B. and {Deluce}, C. and {Depatie}, M.~M. and {Descamps}, F. and {Di Lodovico}, F. and {Dittmer}, J. and {Doxtator}, A. and {Duhaime}, N. and {Duncan}, F. and {Dunger}, J. and {Earle}, A.~D. and {Fabris}, D. and {Falk}, E. and {Farrugia}, A. and {Fatemighomi}, N. and {Felber}, C. and {Fischer}, V. and {Fletcher}, E. and {Ford}, R. and {Frankiewicz}, K. and {Gagnon}, N. and {Gaur}, A. and {Gauthier}, J. and {Gibson-Foster}, A. and {Gilje}, K. and {Gonz{\'a}lez-Reina}, O.~I. and {Gooding}, D. and {Gorel}, P. and {Graham}, K. and {Grant}, C. and {Grove}, J. and {Grullon}, S. and {Guillian}, E. and {Hall}, S. and {Hallin}, A.~L. and {Hallman}, D. and {Hans}, S. and {Hartnell}, J. and {Harvey}, P. and {Hedayatipour}, M. and {Heintzelman}, W.~J. and {Heise}, J. and {Helmer}, R.~L. and {Hodak}, B. and {Hodak}, M. and {Hood}, M. and {Horne}, D. and {Hreljac}, B. and {Hu}, J. and {Hussain}, S.~M.~A. and {Iida}, T. and {In{\'a}cio}, A.~S. and {Jackson}, C.~M. and {Jelley}, N.~A. and {Jillings}, C.~J. and {Jones}, C. and {Jones}, P.~G. and {Kamdin}, K. and {Kaptanoglu}, T. and {Kaspar}, J. and {Keeter}, K. and {Kefelian}, C. and {Khaghani}, P. and {Kippenbrock}, L. and {Klein}, J.~R. and {Knapik}, R. and {Kofron}, J. and {Kormos}, L.~L. and {Korte}, S. and {Krar}, B. and {Kraus}, C. and {Krauss}, C.~B. and {Kroupov{\'a}}, T. and {Labe}, K. and {Lafleur}, F. and {Lam}, I. and {Lan}, C. and {Land}, B.~J. and {Lane}, R. and {Langrock}, S. and {Larochelle}, P. and {Larose}, S. and {LaTorre}, A. and {Lawson}, I. and {Lebanowski}, L. and {Lefeuvre}, G.~M. and {Leming}, E.~J. and {Li}, A. and {Li}, O. and {Lidgard}, J. and {Liggins}, B. and {Liimatainen}, P. and {Lin}, Y.~H. and {Liu}, X. and {Liu}, Y. and {Lozza}, V. and {Luo}, M. and {Maguire}, S. and {Maio}, A. and {Majumdar}, K. and {Manecki}, S. and {Maneira}, J. and {Martin}, R.~D. and {Marzec}, E. and {Mastbaum}, A. and {Mathewson}, A. and {McCauley}, N. and {McDonald}, A.~B. and {McFarlane}, K. and {Mekarski}, P. and {Meyer}, M. and {Miller}, C. and {Mills}, C. and {Mlejnek}, M. and {Mony}, E. and {Morissette}, B. and {Morton-Blake}, I. and {Mottram}, M.~J. and {Nae}, S. and {Nirkko}, M. and {Nolan}, L.~J. and {Novikov}, V.~M. and {O'Keeffe}, H.~M. and {O'Sullivan}, E. and {Orebi Gann}, G.~D. and {Parnell}, M.~J. and {Paton}, J. and {Peeters}, S.~J.~M. and {Pershing}, T. and {Petriw}, Z. and {Petzoldt}, J. and {Pickard}, L. and {Pracsovics}, D. and {Prior}, G. and {Prouty}, J.~C. and {Quirk}, S. and {Read}, S. and {Reichold}, A. and {Riccetto}, S. and {Richardson}, R. and {Rigan}, M. and {Ritchie}, I. and {Robertson}, A. and {Robertson}, B.~C. and {Rose}, J. and {Rosero}, R. and {Rost}, P.~M. and {Rumleskie}, J. and {Schumaker}, M.~A. and {Schwendener}, M.~H. and {Scislowski}, D. and {Secrest}, J. and {Seddighin}, M. and {Segui}, L. and {Seibert}, S.},
        title = "{The SNO+ Experiment}",
      journal = {arXiv e-prints},
         year = 2021,
        month = apr,
          eid = {arXiv:2104.11687},
        pages = {arXiv:2104.11687},
          doi = {10.48550/arXiv.2104.11687},
archivePrefix = {arXiv},
       eprint = {2104.11687},
 primaryClass = {physics.ins-det},
       adsurl = {https://ui.adsabs.harvard.edu/abs/2021arXiv210411687S}
}

@ARTICLE{JUNO_preSN,
       author = {{Abusleme}, Angel and {Adam}, Thomas and {Ahmad}, Shakeel and {Ahmed}, Rizwan and {Aiello}, Sebastiano and {Akram}, Muhammad and {Aleem}, Abid and {An}, Fengpeng and {An}, Qi and {Andronico}, Giuseppe and {Anfimov}, Nikolay and {Antonelli}, Vito and {Antoshkina}, Tatiana and {Asavapibhop}, Burin and {de Andr{\'e}}, Jo{\~a}o Pedro Athayde Marcondes and {Auguste}, Didier and {Bai}, Weidong and {Balashov}, Nikita and {Baldini}, Wander and {Barresi}, Andrea and {Basilico}, Davide and {Baussan}, Eric and {Bellato}, Marco and {Beretta}, Marco and {Bergnoli}, Antonio and {Bick}, Daniel and {Bieger}, Lukas and {Biktemerova}, Svetlana and {Birkenfeld}, Thilo and {Morton-Blake}, Iwan and {Blum}, David and {Blyth}, Simon and {Bolshakova}, Anastasia and {Bongrand}, Mathieu and {Bordereau}, Cl{\'e}ment and {Breton}, Dominique and {Brigatti}, Augusto and {Brugnera}, Riccardo and {Bruno}, Riccardo and {Budano}, Antonio and {Busto}, Jose and {Cabrera}, Anatael and {Caccianiga}, Barbara and {Cai}, Hao and {Cai}, Xiao and {Cai}, Yanke and {Cai}, Zhiyan and {Callier}, St{\'e}phane and {Cammi}, Antonio and {Campeny}, Agustin and {Cao}, Chuanya and {Cao}, Guofu and {Cao}, Jun and {Caruso}, Rossella and {Cerna}, C{\'e}dric and {Cerrone}, Vanessa and {Chan}, Chi and {Chang}, Jinfan and {Chang}, Yun and {Chatrabhuti}, Auttakit and {Chen}, Chao and {Chen}, Guoming and {Chen}, Pingping and {Chen}, Shaomin and {Chen}, Yixue and {Chen}, Yu and {Chen}, Zhangming and {Chen}, Zhiyuan and {Chen}, Zikang and {Cheng}, Jie and {Cheng}, Yaping and {Cheng}, Yu Chin and {Chepurnov}, Alexander and {Chetverikov}, Alexey and {Chiesa}, Davide and {Chimenti}, Pietro and {Chin}, Yen-Ting and {Chu}, Ziliang and {Chukanov}, Artem and {Claverie}, G{\'e}rard and {Clementi}, Catia and {Clerbaux}, Barbara and {Molla}, Marta Colomer and {Lorenzo}, Selma Conforti Di and {Coppi}, Alberto and {Corti}, Daniele and {Csakli}, Simon and {Corso}, Flavio Dal and {Dalager}, Olivia and {Datta}, Jaydeep and {Taille}, Christophe De La and {Deng}, Zhi and {Deng}, Ziyan and {Ding}, Xiaoyu and {Ding}, Xuefeng and {Ding}, Yayun and {Dirgantara}, Bayu and {Dittrich}, Carsten and {Dmitrievsky}, Sergey and {Dohnal}, Tadeas and {Dolzhikov}, Dmitry and {Donchenko}, Georgy and {Dong}, Jianmeng and {Doroshkevich}, Evgeny and {Dou}, Wei and {Dracos}, Marcos and {Druillole}, Fr{\'e}d{\'e}ric and {Du}, Ran and {Du}, Shuxian and {Dugas}, Katherine and {Dusini}, Stefano and {Duyang}, Hongyue and {Eck}, Jessica and {Enqvist}, Timo and {Fabbri}, Andrea and {Fahrendholz}, Ulrike and {Fan}, Lei and {Fang}, Jian and {Fang}, Wenxing and {Fargetta}, Marco and {Fedoseev}, Dmitry and {Fei}, Zhengyong and {Feng}, Li-Cheng and {Feng}, Qichun and {Ferraro}, Federico and {Fournier}, Am{\'e}lie and {Gan}, Haonan and {Gao}, Feng and {Garfagnini}, Alberto and {Gavrikov}, Arsenii and {Giammarchi}, Marco and {Giudice}, Nunzio and {Gonchar}, Maxim and {Gong}, Guanghua and {Gong}, Hui and {Gornushkin}, Yuri and {G{\"o}ttel}, Alexandre and {Grassi}, Marco and {Gromov}, Maxim and {Gromov}, Vasily and {Gu}, Minghao and {Gu}, Xiaofei and {Gu}, Yu and {Guan}, Mengyun and {Guan}, Yuduo and {Guardone}, Nunzio and {Guo}, Cong and {Guo}, Wanlei and {Guo}, Xinheng and {Hagner}, Caren and {Han}, Ran and {Han}, Yang and {He}, Miao and {He}, Wei and {Heinz}, Tobias and {Hellmuth}, Patrick and {Heng}, Yuekun and {Herrera}, Rafael and {Hor}, Yuenkeung and {Hou}, Shaojing and {Hsiung}, Yee and {Hu}, Bei-Zhen and {Hu}, Hang and {Hu}, Jianrun and {Hu}, Jun and {Hu}, Shouyang and {Hu}, Tao and {Hu}, Yuxiang and {Hu}, Zhuojun and {Huang}, Guihong and {Huang}, Hanxiong and {Huang}, Jinhao and {Huang}, Junting and {Huang}, Kaixuan and {Huang}, Wenhao and {Huang}, Xin and {Huang}, Xingtao and {Huang}, Yongbo and {Hui}, Jiaqi and {Huo}, Lei and {Huo}, Wenju and {Huss}, C{\'e}dric and {Hussain}, Safeer and {Imbert}, Leonard and {Ioannisian}, Ara and {Isocrate}, Roberto and {Jafar}, Arshak and {Jelmini}, Beatrice and {Jeria}, Ignacio and {Ji}, Xiaolu and {Jia}, Huihui and {Jia}, Junji and {Jian}, Siyu and {Jiang}, Cailian and {Jiang}, Di and {Jiang}, Wei and {Jiang}, Xiaoshan and {Jing}, Xiaoping and {Jollet}, C{\'e}cile and {Kampmann}, Philipp},
        title = "{Real-time monitoring for the next core-collapse supernova in JUNO}",
      journal = {\jcap},
         year = 2024,
        month = jan,
       volume = {2024},
       number = {1},
          eid = {057},
        pages = {057},
          doi = {10.1088/1475-7516/2024/01/057},
archivePrefix = {arXiv},
       eprint = {2309.07109},
 primaryClass = {hep-ex},
       adsurl = {https://ui.adsabs.harvard.edu/abs/2024JCAP...01..057A}
}

@ARTICLE{HyperK,
       author ={{Abe}, K. and {Abe}, Ke. and {Aihara}, H. and {Aimi}, A. and {Akutsu}, R. and {Andreopoulos}, C. and {Anghel}, I. and {Anthony}, L.~H.~V. and {Antonova}, M. and {Ashida}, Y. and {Aushev}, V. and {Barbi}, M. and {Barker}, G.~J. and {Barr}, G. and {Beltrame}, P. and {Berardi}, V. and {Bergevin}, M. and {Berkman}, S. and {Berns}, L. and {Berry}, T. and {Bhadra}, S. and {Bravo-Bergu{\~n}o}, D. and {Blaszczyk}, F. d. M. and {Blondel}, A. and {Bolognesi}, S. and {Boyd}, S.~B. and {Bravar}, A. and {Bronner}, C. and {Buizza Avanzini}, M. and {Cafagna}, F.~S. and {Cole}, A. and {Calland}, R. and {Cao}, S. and {Cartwright}, S.~L. and {Catanesi}, M.~G. and {Checchia}, C. and {Chen-Wishart}, Z. and {Choi}, J.~H. and {Choi}, K. and {Coleman}, J. and {Collazuol}, G. and {Cowan}, G. and {Cremonesi}, L. and {Dealtry}, T. and {De Rosa}, G. and {Densham}, C. and {Dewhurst}, D. and {Drakopoulou}, E.~L. and {Di Lodovico}, F. and {Drapier}, O. and {Dumarchez}, J. and {Dunne}, P. and {Dziewiecki}, M. and {Emery}, S. and {Esmaili}, A. and {Evangelisti}, A. and {Fernandez-Martinez}, E. and {Feusels}, T. and {Finch}, A. and {Fiorentini}, A. and {Fiorillo}, G. and {Fitton}, M. and {Frankiewicz}, K. and {Friend}, M. and {Fujii}, Y. and {Fukuda}, Y. and {Fukuda}, D. and {Ganezer}, K. and {Giganti}, C. and {Gonin}, M. and {Grant}, N. and {Gumplinger}, P. and {Hadley}, D.~R. and {Hartfiel}, B. and {Hartz}, M. and {Hayato}, Y. and {Hayrapetyan}, K. and {Hill}, J. and {Hirota}, S. and {Horiuchi}, S. and {Ichikawa}, A.~K. and {Iijima}, T. and {Ikeda}, M. and {Imber}, J. and {Inoue}, K. and {Insler}, J. and {Intonti}, R.~A. and {Ioannisian}, A. and {Ishida}, T. and {Ishino}, H. and {Ishitsuka}, M. and {Itow}, Y. and {Iwamoto}, K. and {Izmaylov}, A. and {Jamieson}, B. and {Jang}, H.~I. and {Jang}, J.~S. and {Jeon}, S.~H. and {Jiang}, M. and {Jonsson}, P. and {Joo}, K.~K. and {Kaboth}, A. and {Kachulis}, C. and {Kajita}, T. and {Kameda}, J. and {Kataoka}, Y. and {Katori}, T. and {Kayrapetyan}, K. and {Kearns}, E. and {Khabibullin}, M. and {Khotjantsev}, A. and {Kim}, J.~H. and {Kim}, J.~Y. and {Kim}, S.~B. and {Kim}, S.~Y. and {King}, S. and {Kishimoto}, Y. and {Kobayashi}, T. and {Koga}, M. and {Konaka}, A. and {Kormos}, L.~L. and {Koshio}, Y. and {Korzenev}, A. and {Kowalik}, K.~L. and {Kropp}, W.~R. and {Kudenko}, Y. and {Kurjata}, R. and {Kutter}, T. and {Kuze}, M. and {Labarga}, L. and {Lagoda}, J. and {Lasorak}, P.~J.~J. and {Laveder}, M. and {Lawe}, M. and {Learned}, J.~G. and {Lim}, I.~T. and {Lindner}, T. and {Litchfield}, R.~P. and {Longhin}, A. and {Loverre}, P. and {Lou}, T. and {Ludovici}, L. and {Ma}, W. and {Magaletti}, L. and {Mahn}, K. and {Malek}, M. and {Maret}, L. and {Mariani}, C. and {Martens}, K. and {Marti}, Ll. and {Martin}, J.~F. and {Marzec}, J. and {Matsuno}, S. and {Mazzucato}, E. and {McCarthy}, M. and {McCauley}, N. and {McFarland}, K.~S. and {McGrew}, C. and {Mefodiev}, A. and {Mermod}, P. and {Metelko}, C. and {Mezzetto}, M. and {Migenda}, J. and {Mijakowski}, P. and {Minakata}, H. and {Minamino}, A. and {Mine}, S. and {Mineev}, O. and {Mitra}, A. and {Miura}, M. and {Mochizuki}, T. and {Monroe}, J. and {Moon}, D.~H. and {Moriyama}, S. and {Mueller}, T. and {Muheim}, F. and {Murase}, K. and {Muto}, F. and {Nakahata}, M. and {Nakajima}, Y. and {Nakamura}, K. and {Nakaya}, T. and {Nakayama}, S. and {Nantais}, C. and {Needham}, M. and {Nicholls}, T. and {Nishimura}, Y. and {Noah}, E. and {Nova}, F. and {Nowak}, J. and {Nunokawa}, H. and {Obayashi}, Y. and {O'Keeffe}, H.~M. and {Okajima}, Y. and {Okumura}, K. and {Onishchuk}, Yu. and {O'Sullivan}, E. and {O'Sullivan}, L.},
        title = "{Hyper-Kamiokande Design Report}",
      journal = {arXiv e-prints},
         year = 2018,
        month = may,
          eid = {arXiv:1805.04163},
        pages = {arXiv:1805.04163},
          doi = {10.48550/arXiv.1805.04163},
archivePrefix = {arXiv},
       eprint = {1805.04163},
 primaryClass = {physics.ins-det},
       adsurl = {https://ui.adsabs.harvard.edu/abs/2018arXiv180504163H}
}

@article{SNEWS,
    author = "Antonioli, Pietro and others",
    title = "{SNEWS: The Supernova Early Warning System}",
    eprint = "astro-ph/0406214",
    archivePrefix = "arXiv",
    doi = "10.1088/1367-2630/6/1/114",
    journal = "New J. Phys.",
    volume = "6",
    pages = "114",
    year = "2004"
}

@ARTICLE{SNEWS_2.0,
       author = {{Al Kharusi}, S. and {BenZvi}, S.~Y. and {Bobowski}, J.~S. and {Bonivento}, W. and {Brdar}, V. and {Brunner}, T. and {Caden}, E. and {Clark}, M. and {Coleiro}, A. and {Colomer-Molla}, M. and {Crespo-Anad{\'o}n}, J.~I. and {Depoian}, A. and {Dornic}, D. and {Fischer}, V. and {Franco}, D. and {Fulgione}, W. and {Gallo Rosso}, A. and {Geske}, M. and {Griswold}, S. and {Gromov}, M. and {Haggard}, D. and {Habig}, A. and {Halim}, O. and {Higuera}, A. and {Hill}, R. and {Horiuchi}, S. and {Ishidoshiro}, K. and {Kato}, C. and {Katsavounidis}, E. and {Khaitan}, D. and {Kneller}, J.~P. and {Kopec}, A. and {Kulikovskiy}, V. and {Lai}, M. and {Lamoureux}, M. and {Lang}, R.~F. and {Li}, H.~L. and {Lincetto}, M. and {Lunardini}, C. and {Migenda}, J. and {Milisavljevic}, D. and {McCarthy}, M.~E. and {O Connor}, E. and {O Sullivan}, E. and {Pagliaroli}, G. and {Patel}, D. and {Peres}, R. and {Pointon}, B.~W. and {Qin}, J. and {Raj}, N. and {Renshaw}, A. and {Roeth}, A. and {Rumleskie}, J. and {Scholberg}, K. and {Sheshukov}, A. and {Sonley}, T. and {Strait}, M. and {Takhistov}, V. and {Tamborra}, I. and {Tseng}, J. and {Tunnell}, C.~D. and {Vasel}, J. and {Vigorito}, C.~F. and {Viren}, B. and {Virtue}, C.~J. and {Wang}, J.~S. and {Wen}, L.~J. and {Winslow}, L. and {Wolfs}, F.~L.~H. and {Xu}, X.~J. and {Xu}, Y.},
        title = "{SNEWS 2.0: a next-generation supernova early warning system for multi-messenger astronomy}",
      journal = {New Journal of Physics},
         year = 2021,
        month = mar,
       volume = {23},
       number = {3},
          eid = {031201},
        pages = {031201},
          doi = {10.1088/1367-2630/abde33},
archivePrefix = {arXiv},
       eprint = {2011.00035},
 primaryClass = {astro-ph.HE},
       adsurl = {https://ui.adsabs.harvard.edu/abs/2021NJPh...23c1201A}
}

@article{odrzywolek2,
    author = "Odrzywolek, A. and Misiaszek, Marcin and Kutschera, M.",
    title = "{Detection possibility of the pair - annihilation neutrinos from the neutrino - cooled pre-supernova star}",
    eprint = "astro-ph/0311012",
    archivePrefix = "arXiv",
    doi = "10.1016/j.astropartphys.2004.02.002",
    journal = "Astropart. Phys.",
    volume = "21",
    pages = "303--313",
    year = "2004"
}

@article{odrzywolek,
    author = "Odrzywolek, Andrzej and Heger, Alexander",
    editor = "Zalewska, Agnieszka",
    title = "{Neutrino signatures of dying massive stars: From main sequence to the neutron star}",
    journal = "Acta Phys. Polon. B",
    volume = "41",
    pages = "1611--1628",
    year = "2010"
}

@article{patton,
    author = "Patton, Kelly M. and Lunardini, Cecilia and Farmer, Robert J. and Timmes, F. X.",
    title = "{Neutrinos from beta processes in a presupernova: probing the isotopic evolution of a massive star}",
    eprint = "1709.01877",
    archivePrefix = "arXiv",
    primaryClass = "astro-ph.HE",
    reportNumber = "INT-PUB-17-037",
    doi = "10.3847/1538-4357/aa95c4",
    journal = "Astrophys. J.",
    volume = "851",
    number = "1",
    pages = "6",
    year = "2017"
}

@article{kato_preSN,
    author = "Kato, Chinami and Nagakura, Hiroki and Furusawa, Shun and Takahashi, Koh and Umeda, Hideyuki and Yoshida, Takashi and Ishidoshiro, Koji and Yamada, Shoichi",
    title = "{Neutrino emissions in all flavors up to the pre-bounce of massive stars and the possibility of their detections}",
    eprint = "1704.05480",
    archivePrefix = "arXiv",
    primaryClass = "astro-ph.HE",
    doi = "10.3847/1538-4357/aa8b72",
    journal = "Astrophys. J.",
    volume = "848",
    number = "1",
    pages = "48",
    year = "2017"
}

@ARTICLE{Kato2015,
       author = {{Kato}, Chinami and {Delfan Azari}, Milad and {Yamada}, Shoichi and {Takahashi}, Koh and {Umeda}, Hideyuki and {Yoshida}, Takashi and {Ishidoshiro}, Koji},
        title = "{Pre-supernova Neutrino Emissions from ONe Cores in the Progenitors of Core-collapse Supernovae: Are They Distinguishable from Those of Fe Cores?}",
      journal = {\apj},
         year = 2015,
        month = aug,
       volume = {808},
       number = {2},
          eid = {168},
        pages = {168},
          doi = {10.1088/0004-637X/808/2/168},
archivePrefix = {arXiv},
       eprint = {1506.02358},
 primaryClass = {astro-ph.HE},
       adsurl = {https://ui.adsabs.harvard.edu/abs/2015ApJ...808..168K}
}

@ARTICLE{Yoshida2016,
       author = {{Yoshida}, Takashi and {Takahashi}, Koh and {Umeda}, Hideyuki and {Ishidoshiro}, Koji},
        title = "{Presupernova neutrino events relating to the final evolution of massive stars}",
      journal = {\prd},
         year = 2016,
        month = jun,
       volume = {93},
       number = {12},
          eid = {123012},
        pages = {123012},
          doi = {10.1103/PhysRevD.93.123012},
archivePrefix = {arXiv},
       eprint = {1606.04915},
 primaryClass = {astro-ph.HE},
       adsurl = {https://ui.adsabs.harvard.edu/abs/2016PhRvD..93l3012Y}
}

@article{MESA_reference,
    author = "Paxton, Bill and Bildsten, Lars and Dotter, Aaron and Herwig, Falk and Lesaffre, Pierre and Timmes, Frank",
    collaboration = "MESA",
    title = "{Modules for Experiments in Stellar Astrophysics (MESA)}",
    eprint = "1009.1622",
    archivePrefix = "arXiv",
    primaryClass = "astro-ph.SR",
    doi = "10.1088/0067-0049/192/1/3",
    journal = "Astrophys. J. Suppl.",
    volume = "192",
    pages = "3",
    year = "2011"
}

@ARTICLE{Kato2020_review,
       author = {{Kato}, C. and {Ishidoshiro}, K. and {Yoshida}, T.},
        title = "{Theoretical Prediction of Presupernova Neutrinos and Their Detection}",
      journal = {Annual Review of Nuclear and Particle Science},
         year = 2020,
        month = oct,
       volume = {70},
        pages = {121-145},
          doi = {10.1146/annurev-nucl-040620-021320},
archivePrefix = {arXiv},
       eprint = {2006.02519},
 primaryClass = {astro-ph.HE},
       adsurl = {https://ui.adsabs.harvard.edu/abs/2020ARNPS..70..121K}
}

@article{KamLAND_detector,
    author = "Suzuki, Atsuto",
    title = "{Antineutrino Science in KamLAND}",
    eprint = "1409.4515",
    archivePrefix = "arXiv",
    primaryClass = "hep-ex",
    doi = "10.1140/epjc/s10052-014-3094-x",
    journal = "Eur. Phys. J. C",
    volume = "74",
    number = "10",
    pages = "3094",
    year = "2014"
}

@article{KamLANDZen,
    author = "Gando, A. and others",
    collaboration = "KamLAND-Zen",
    title = "{Measurement of the double-$\beta$ decay half-life of $^{136}Xe$ with the KamLAND-Zen experiment}",
    eprint = "1201.4664",
    archivePrefix = "arXiv",
    primaryClass = "hep-ex",
    doi = "10.1103/PhysRevC.85.045504",
    journal = "Phys. Rev. C",
    volume = "85",
    pages = "045504",
    year = "2012"
}

@article{KamLANDZen2025,
  title = {Search for Majorana Neutrinos with the Complete KamLAND-Zen Dataset},
  author = {Abe, S. and Araki, T. and Chiba, K. and Eda, T. and Eizuka, M. and Funahashi, Y. and Furuto, A. and Gando, A. and Gando, Y. and Goto, S. and Hachiya, T. and Hata, K. and Ichimura, K. and Ieki, S. and Ikeda, H. and Inoue, K. and Ishidoshiro, K. and Kamei, Y. and Kawada, N. and Kishimoto, Y. and Koga, M. and Marthe, A. and Matsumoto, Y. and Mitsui, T. and Miyake, H. and Morita, D. and Nakajima, R. and Nakamura, K. and Nakamura, R. and Nakamura, R. and Nakane, J. and Ono, T. and Ozaki, H. and Saito, K. and Sakai, T. and Shimizu, I. and Shirai, J. and Shiraishi, K. and Suzuki, A. and Tachibana, K. and Tamae, K. and Watanabe, H. and Watanabe, K. and Yoshida, S. and Umehara, S. and Fushimi, K. and Kotera, K. and Urano, Y. and Berger, B. E. and Fujikawa, B. K. and Learned, J. G. and Maricic, J. and Fu, Z. and Ghosh, S. and Smolsky, J. and Winslow, L. A. and Efremenko, Y. and Karwowski, H. J. and Markoff, D. M. and Tornow, W. and Dell'Oro, S. and O'Donnell, T. and Detwiler, J. A. and Enomoto, S. and Decowski, M. P. and Weerman, K. M. and Grant, C. and Penek, \"O. and Song, H. and Li, A. and Axani, S. N. and Garcia, M. and Sarfraz, M.},
  collaboration = {KamLAND-Zen Collaboration},
  journal = {Phys. Rev. Lett.},
  volume = {135},
  issue = {26},
  pages = {262501},
  numpages = {7},
  year = {2025},
  month = {Dec},
  publisher = {American Physical Society},
  doi = {10.1103/jkf6-48j8},
  url = {https://link.aps.org/doi/10.1103/jkf6-48j8}
}

@article{KamLAND_alphan,
    author = "Abe, S. and others",
    collaboration = "KamLAND",
    title = "{Precision Measurement of Neutrino Oscillation Parameters with KamLAND}",
    eprint = "0801.4589",
    archivePrefix = "arXiv",
    primaryClass = "hep-ex",
    doi = "10.1103/PhysRevLett.100.221803",
    journal = "Phys. Rev. Lett.",
    volume = "100",
    pages = "221803",
    year = "2008"
}

@article{kamland_preSN,
    author = "Asakura, K. and others",
    collaboration = "KamLAND",
    title = "{KamLAND Sensitivity to Neutrinos from Pre-Supernova Stars}",
    eprint = "1506.01175",
    archivePrefix = "arXiv",
    primaryClass = "astro-ph.HE",
    doi = "10.3847/0004-637X/818/1/91",
    journal = "Astrophys. J.",
    volume = "818",
    number = "1",
    pages = "91",
    year = "2016"
}

@article{Super-Kamiokande:2002weg,
    author = "Fukuda, Y. and others",
    editor = "Ilyin, V. A. and Korenkov, V. V. and Perret-Gallix, D.",
    collaboration = "Super-Kamiokande",
    title = "{The Super-Kamiokande detector}",
    doi = "10.1016/S0168-9002(03)00425-X",
    journal = "Nucl. Instrum. Meth. A",
    volume = "501",
    pages = "418--462",
    year = "2003"
}

@article{Super-Kamiokande:2021the,
    author = "Abe, K. and others",
    collaboration = "Super-Kamiokande",
    title = "{First gadolinium loading to Super-Kamiokande}",
    eprint = "2109.00360",
    archivePrefix = "arXiv",
    primaryClass = "physics.ins-det",
    doi = "10.1016/j.nima.2021.166248",
    journal = "Nucl. Instrum. Meth. A",
    volume = "1027",
    pages = "166248",
    year = "2022"
}

@ARTICLE{2024NIMPA106569480A,
       author = {{Abe}, K. and {Bronner}, C. and {Hayato}, Y. and {Hiraide}, K. and {Hosokawa}, K. and {Ieki}, K. and {Ikeda}, M. and {Kameda}, J. and {Kanemura}, Y. and {Kaneshima}, R. and {Kashiwagi}, Y. and {Kataoka}, Y. and {Miki}, S. and {Mine}, S. and {Miura}, M. and {Moriyama}, S. and {Nakano}, Y. and {Nakahata}, M. and {Nakayama}, S. and {Noguchi}, Y. and {Sato}, K. and {Sekiya}, H. and {Shiba}, H. and {Shimizu}, K. and {Shiozawa}, M. and {Sonoda}, Y. and {Suzuki}, Y. and {Takeda}, A. and {Takemoto}, Y. and {Tanaka}, H. and {Yano}, T. and {Han}, S. and {Kajita}, T. and {Okumura}, K. and {Tashiro}, T. and {Tomiya}, T. and {Wang}, X. and {Yoshida}, S. and {Fernandez}, P. and {Labarga}, L. and {Ospina}, N. and {Zaldivar}, B. and {Pointon}, B.~W. and {Kearns}, E. and {Raaf}, J.~L. and {Wan}, L. and {Wester}, T. and {Bian}, J. and {Griskevich}, N.~J. and {Smy}, M.~B. and {Sobel}, H.~W. and {Takhistov}, V. and {Yankelevich}, A. and {Hill}, J. and {Jang}, M.~C. and {Lee}, S.~H. and {Moon}, D.~H. and {Park}, R.~G. and {Bodur}, B. and {Scholberg}, K. and {Walter}, C.~W. and {Beauch{\^e}ne}, A. and {Drapier}, O. and {Giampaolo}, A. and {Mueller}, Th. A. and {Santos}, A.~D. and {Paganini}, P. and {Quilain}, B. and {Rogly}, R. and {Nakamura}, T. and {Jang}, J.~S. and {Machado}, L.~N. and {Learned}, J.~G. and {Choi}, K. and {Iovine}, N. and {Cao}, S. and {Anthony}, L.~H.~V. and {Martin}, D. and {Prouse}, N.~W. and {Scott}, M. and {Uchida}, Y. and {Berardi}, V. and {Calabria}, N.~F. and {Catanesi}, M.~G. and {Radicioni}, E. and {Langella}, A. and {De Rosa}, G. and {Collazuol}, G. and {Iacob}, F. and {Mattiazzi}, M. and {Ludovici}, L. and {Gonin}, M. and {P{\'e}riss{\'e}}, L. and {Pronost}, G. and {Fujisawa}, C. and {Maekawa}, Y. and {Nishimura}, Y. and {Okazaki}, R. and {Akutsu}, R. and {Friend}, M. and {Hasegawa}, T. and {Ishida}, T. and {Kobayashi}, T. and {Jakkapu}, M. and {Matsubara}, T. and {Nakadaira}, T. and {Nakamura}, K. and {Oyama}, Y. and {Sakashita}, K. and {Sekiguchi}, T. and {Tsukamoto}, T. and {Bhuiyan}, N. and {Burton}, G.~T. and {Di Lodovico}, F. and {Gao}, J. and {Goldsack}, A. and {Katori}, T. and {Migenda}, J. and {Ramsden}, R.~M. and {Xie}, Z. and {Zsoldos}, S. and {Suzuki}, A.~T. and {Takagi}, Y. and {Takeuchi}, Y. and {Zhong}, H. and {Feng}, J. and {Feng}, L. and {Hu}, J.~R. and {Hu}, Z. and {Kawaue}, M. and {Kikawa}, T. and {Mori}, M. and {Nakaya}, T. and {Wendell}, R.~A. and {Yasutome}, K. and {Jenkins}, S.~J. and {McCauley}, N. and {Mehta}, P. and {Tarant}, A. and {Wilking}, M.~J. and {Fukuda}, Y. and {Itow}, Y. and {Menjo}, H. and {Ninomiya}, K. and {Yoshioka}, Y. and {Lagoda}, J. and {Mandal}, M. and {Mijakowski}, P. and {Prabhu}, Y.~S. and {Zalipska}, J. and {Jia}, M. and {Jiang}, J. and {Shi}, W. and {Yanagisawa}, C. and {Harada}, M. and {Hino}, Y. and {Ishino}, H. and {Koshio}, Y. and {Nakanishi}, F. and {Sakai}, S. and {Tada}, T. and {Tano}, T. and {Ishizuka}, T. and {Barr}, G. and {Barrow}, D. and {Cook}, L. and {Samani}, S. and {Wark}, D. and {Holin}, A. and {Nova}, F. and {Jung}, S. and {Yang}, B.~S. and {Yang}, J.~Y. and {Yoo}, J. and {Fannon}, J.~E.~P. and {Kneale}, L. and {Malek}, M. and {McElwee}, J.~M. and {Thiesse}, M.~D. and {Thompson}, L.~F. and {Wilson}, S.~T. and {Okazawa}, H. and {Lakshmi}, S.~M. and {Kim}, S.~B. and {Kwon}, E. and {Seo}, J.~W. and {Yu}, I. and {Ichikawa}, A.~K. and {Nakamura}, K. and {Tairafune}, S. and {Nishijima}, K. and {Eguchi}, A. and {Nakagiri}, K. and {Nakajima}, Y. and {Shima}, S. and {Taniuchi}, N. and {Watanabe}, E. and {Yokoyama}, M. and {de Perio}, P. and {Fujita}, S.},
        title = "{Second gadolinium loading to Super-Kamiokande}",
      journal = {Nuclear Instruments and Methods in Physics Research A},
         year = 2024,
        month = aug,
       volume = {1065},
          eid = {169480},
        pages = {169480},
          doi = {10.1016/j.nima.2024.169480},
archivePrefix = {arXiv},
       eprint = {2403.07796},
 primaryClass = {physics.ins-det},
       adsurl = {https://ui.adsabs.harvard.edu/abs/2024NIMPA106569480A}
}

@article{lucaspaper,
    author = "Machado, L. N. and others",
    collaboration = "Super-Kamiokande",
    title = "{Pre-supernova Alert System for Super-Kamiokande}",
    eprint = "2205.09881",
    archivePrefix = "arXiv",
    primaryClass = "hep-ex",
    doi = "10.3847/1538-4357/ac7f9c",
    journal = "Astrophys. J.",
    volume = "935",
    number = "1",
    pages = "40",
    year = "2022"
}

@article{Huber2011,
  author = "Huber, Patrick",
  title = "{Determination of antineutrino spectra from nuclear reactors}",
  doi = "10.1103/PhysRevC.84.024617",
  journal = "Phys. Rev. C",
  volume = "84",
  pages = "024617",
  year = "2011",
}

@article{Mueller2011,
  author = "Mueller, Th. A. and Lhuillier, D. and Fallot, M. and Letourneau, A. and Cormon, S. and Fechner, M. and Giot, L. and Lasserre, T. and Martino, J. and Mention, G. and Porta, A. and Yermia, F.",
  title = "{Improved predictions of reactor antineutrino spectra}",
  doi = "10.1103/PhysRevC.83.054615",
  journal = "Phys. Rev. C",
  volume = "83",
  pages = "054615",
  year = "2011",
}

@article{Vogel1981,
  author = "Vogel, P. and Schenter, G. K. and Mann, F. M. and Schenter, R. E.",
  title = "{Reactor antineutrino spectra and their application to antineutrino-induced reactions. II}",
  doi = "10.1103/PhysRevC.24.1543",
  journal = "Phys. Rev. C",
  volume = "24",
  pages = "1543--1553",
  year = "1981",
}

@article{Enomoto_geo,
author = "S. Enomoto and E. Ohtani and K. Inoue and A. Suzuki",
title = "{Neutrino geophysics with KamLAND and future prospects}",
doi = "https://doi.org/10.1016/j.epsl.2007.03.038",
journal = "Earth and Planetary Science Letters",
volume = "258",
pages = "147-159",
year = "2007",
}

@article{KamLAND:2013rgu,
    author = "Gando, A. and others",
    collaboration = "KamLAND",
    title = "{Reactor On-Off Antineutrino Measurement with KamLAND}",
    eprint = "1303.4667",
    archivePrefix = "arXiv",
    primaryClass = "hep-ex",
    doi = "10.1103/PhysRevD.88.033001",
    journal = "Phys. Rev. D",
    volume = "88",
    number = "3",
    pages = "033001",
    year = "2013"
}
\end{document}